# BIG BANG NUCLEOSYNTHESIS, IMPLICATIONS OF RECENT CMB DATA AND SUPERSYMMETRIC DARK MATTER[*]

KEITH A. OLIVE

*Theoretical Physics Institute, School of Physics and Astronomy,
University of Minnesota, Minneapolis, MN 55455, USA
E-mail: olive@umn.edu*

The BBN predictions for the abundances of the light element isotopes is reviewed and compared with recent observational data. The single unknown parameter of standard BBN is the baryon-to-photon ratio, $\eta$, and can be determined by the concordance between theory and observation. Recent CMB anisotropy measurements also lead to a determination of $\eta$ and these results are contrasted with those from BBN. In addition, the CMB data indicate that the Universe is spatially flat. Thus it is clear that some form of non-baryonic dark matter or dark energy is necessary. Here I will also review the current expectations for cold dark matter from minimal supersymmetric models. The viability of detecting supersymmetric dark matter will also be discussed.

## 1 Introduction

The cornerstones of the Big Bang theory are the cosmic microwave background radiation (CMB) and big bang nucleosynthesis (BBN) (one could make an argument to include inflation as well). The existence of the former and the success of the latter point unequivocally to a hot and dense origin to the Universe. Indeed, these two aspects of the theory are intimately linked as early work by Alpher and Herman[1] on BBN led to their prediction of the CMB with a temperature of order 10K.

These lectures will focus on recent developments in BBN theory and the related observations which test for concordance. Recent measurements of the CMB power spectrum have allowed an unprecedented level of accuracy in the determination of cosmological parameters including the baryon density, which is the key parameter for BBN. The concordance between BBN and the recent CMB measurements will be addressed. It is also becoming clear that while the total density of the Universe is near critical, i.e., we live in a spatially flat Universe, the baryon density is only a small fraction of the total energy density. Something is missing. While much of missing energy appears to be best fit by a smooth component such as a cosmological constant, a sizable fraction must lie in the form of non-baryonic dark matter. The second half of these lectures will focus on the specific possibility of supersymmetric dark matter.

---

[*]SUMMARY OF INVITED LECTURES AT THE FIRST NCTS WORKSHOP ON ASTROPARTICLE PHYSICS, DECEMBER 6 - 9, 2001, KENTING, TAIWAN.

## 2 Big Bang Nucleosynthesis

The standard model[2] of big bang nucleosynthesis (BBN) is based on the relatively simple idea of including an extended nuclear network into a homogeneous and isotropic cosmology. Apart from the input nuclear cross sections, the theory contains only a single parameter, namely the baryon-to-photon ratio, $\eta$. Other factors, such as the uncertainties in reaction rates, and the neutron mean-life can be treated by standard statistical and Monte Carlo techniques[3]. The theory then allows one to make predictions (with well-defined uncertainties) of the abundances of the light elements, D, $^3$He, $^4$He, and $^7$Li.

### 2.1 Theory

Conditions for the synthesis of the light elements were attained in the early Universe at temperatures $T \gtrsim 1$ MeV. In the early Universe, the energy density was dominated by radiation with

$$\rho = \frac{\pi^2}{30}(2 + \frac{7}{2} + \frac{7}{4}N_\nu)T^4 \quad (1)$$

from the contributions of photons, electrons and positrons, and $N_\nu$ neutrino flavors (at higher temperatures, other particle degrees of freedom should be included as well). At these temperatures, weak interaction rates were in equilibrium. In particular, the processes

$$n + e^+ \leftrightarrow p + \bar{\nu}_e$$
$$n + \nu_e \leftrightarrow p + e^-$$
$$n \leftrightarrow p + e^- + \bar{\nu}_e \quad (2)$$

fix the ratio of number densities of neutrons to protons. At $T \gg 1$ MeV, $(n/p) \simeq 1$.

The weak interactions do not remain in equilibrium at lower temperatures. Freeze-out occurs when the weak interaction rate, $\Gamma_{wk} \sim G_F^2 T^5$ falls below the expansion rate which is given by the Hubble parameter, $H \sim \sqrt{G_N \rho} \sim T^2/M_P$, where $M_P = 1/\sqrt{G_N} \simeq 1.2 \times 10^{19}$ GeV. The $\beta$-interactions in eq. (2) freeze-out at about 0.8 MeV. As the temperature falls and approaches the point where the weak interaction rates are no longer fast enough to maintain equilibrium, the neutron to proton ratio is given approximately by the Boltzmann factor, $(n/p) \simeq e^{-\Delta m/T} \sim 1/6$, where $\Delta m$ is the neutron-proton mass difference. After freeze-out, free neutron decays drop the ratio slightly to about 1/7 before nucleosynthesis begins.

The nucleosynthesis chain begins with the formation of deuterium by the process, $p + n \to D + \gamma$. However, because of the large number of photons relative to nucleons, $\eta^{-1} = n_\gamma/n_B \sim 10^{10}$, deuterium production is delayed past the point where the temperature has fallen below the deuterium binding energy, $E_B = 2.2$ MeV (the average photon energy in a blackbody is $\bar{E}_\gamma \simeq 2.7T$). This is because there are many photons in the exponential tail of the photon energy distribution with energies $E > E_B$ despite the fact that the temperature or $\bar{E}_\gamma$ is less than $E_B$. The degree to which deuterium production is delayed can be found by comparing the qualitative expressions for the deuterium production and destruction rates,

$$\Gamma_p \approx n_B \sigma v \qquad (3)$$
$$\Gamma_d \approx n_\gamma \sigma v e^{-E_B/T}$$

When the quantity $\eta^{-1}\exp(-E_B/T) \sim 1$, the rate for deuterium destruction $(D + \gamma \to p + n)$ finally falls below the deuterium production rate and the nuclear chain begins at a temperature $T \sim 0.1 MeV$.

In addition to the $p\,(n,\gamma)\,D$ reaction, the other major reactions leading to the production of the light elements are:

D (D, p) T    D (n, γ) T    $^3$He (n, p) T

D (D, n) $^3$He    D (p, γ) $^3$He

Followed by the reactions producing $^4$He:

D (D, γ) $^4$He    $^3$He ($^3$He, 2p) $^4$He

D ($^3$He, p) $^4$He    T (p, γ) $^4$He

T (D, n) $^4$He    $^3$He (n, γ) $^4$He

The gap at $A = 5$ is overcome and the production of $^7$Li proceeds through:

$^3$He ($^4$He,γ) $^7$Be

$$\to {}^7\text{Li} + e^+ + \nu_e$$

T ($^4$He,γ) $^7$Li

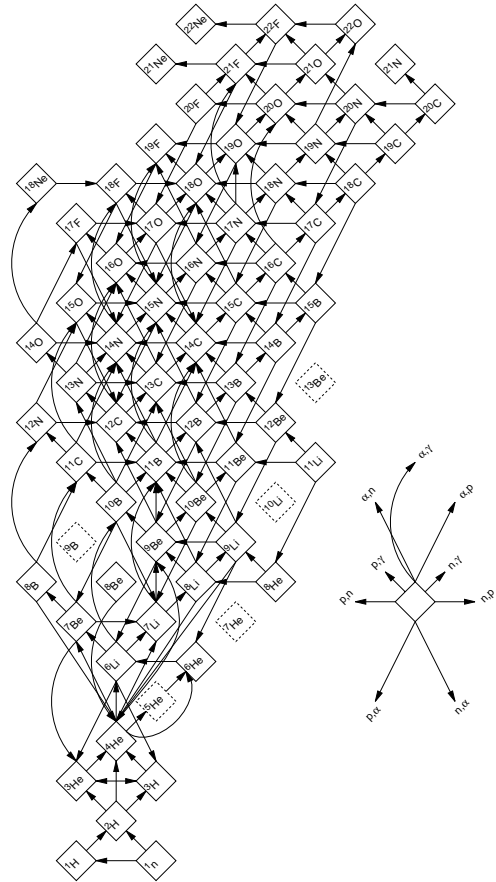

Figure 1. The nuclear network used in BBN calculations.

The gap at $A = 8$ prevents the production of other isotopes in any significant quantity. The nuclear chain in BBN calculations was extended[4] and is shown in Figure 1.

The dominant product of big bang nucleosynthesis is $^4$He and its abun-dance is very sensitive to the $(n/p)$ ratio

$$Y_p = \frac{2(n/p)}{[1 + (n/p)]} \approx 0.25 \quad (4)$$

i.e., an abundance of close to 25% by mass. Lesser amounts of the other light elements are produced: D and $^3$He at the level of about $10^{-5}$ by number, and $^7$Li at the level of $10^{-10}$ by number.

Historically, BBN as a theory explaining the observed element abundances was nearly abandoned due its inability to explain *all* element abundances. Subsequently, stellar nucleosynthesis became the leading theory for element production[5]. However, two key questions persisted. 1) The abundance of $^4$He as a function of metallicity is nearly flat and no abundances are observed to be below about 23% as exaggerated in Fig. 2. In particular, even in systems in which an element such as Oxygen, which traces stellar activity, is observed at extremely low values (compared with the solar value of O/H $\approx 8.5 \times 10^{-4}$), the $^4$He abundance is nearly constant. This is very different from all other element abundances (with the exception of $^7$Li as we will see below). For example, in Figure 3, the N/H vs. O/H correlation is shown. As one can clearly see, the abundance of N/H goes to 0, as O/H goes to 0, indicating a stellar source for Nitrogen. 2) Stellar sources can not produce the observed abundance of D/H. Indeed, stars destroy deuterium and no astrophysical site is known for the production of significant amounts of deuterium[6]. Thus we are led back to BBN for the origins of D, $^3$He, $^4$He, and $^7$Li.

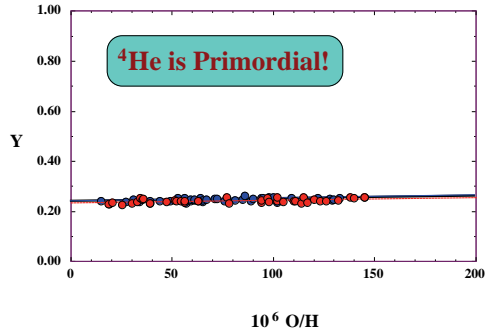

Figure 2. The $^4$He mass fraction as determined in extragalactic H II regions as a function of O/H.

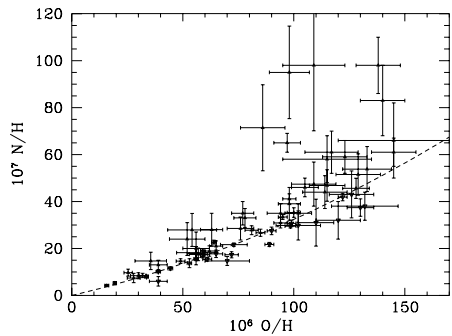

Figure 3. The Nitrogen and Oxygen abundances in the same extragalactic HII regions with observed $^4$He shown in Figure 2.

The resulting abundances of the light elements are shown in Figure 4, over the range in $\eta_{10} = 10^{10}\eta$ between 1 and 10. The left plot shows the abundance of $^4$He by mass, $Y$, and the abundances of the other three isotopes by number. The curves indicate the central predictions from BBN, while the bands correspond to the uncertainty in the predicted abundances based primarily the uncertainty in the input nuclear reactions as computed by Monte Carlo in ref. [7]. This theoretical uncertainty is shown explicitly in the right panel as a function of $\eta_{10}$. The dark shaded boxes correspond to the observed abundances of $^4$He and $^7$Li and will be discussed below. The dashed boxes correspond to the ranges of the elements consistent with the systematic uncertainties in the observations. The broad band shows a liberal range for $\eta_{10}$ consistent with the observations.

At present, there is a general concordance between the theoretical predictions and the observational data, particularly, for $^4$He and $^7$Li. These two elements indicate that $\eta$ lies in the range $1.7 < \eta < 4.7$, corresponding to a range in $\Omega_B h^2 = 0.006 - 0.017^a$. There is limited agreement for D/H as well, as will be discussed below. D/H is compatible with $^4$He and $^7$Li at the $2\sigma$ level in the range $4.7 < \eta < 6.2$ ($\Omega_B h^2 = 0.017 - 0.023$).

---

$^a\Omega$ is the total density of matter relative to the critical density and $\Omega_B$ is the fraction of critical density in baryons. $h$ is Hubble parameter scaled to 100 km Mpc$^{-1}$ s$^{-1}$.

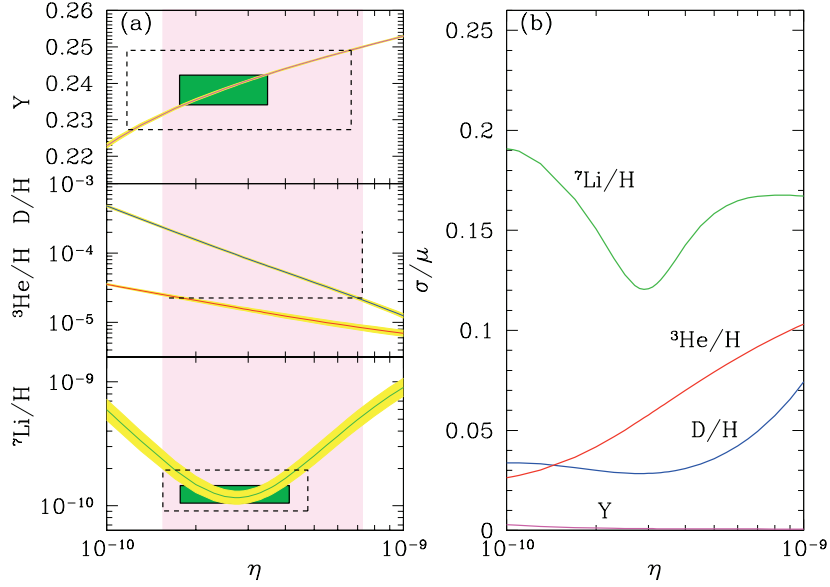

Figure 4. The light element abundances from big bang nucleosynthesis as a function of $\eta_{10}$.

## 2.2 Data-$^4$He

The primordial $^4$He abundance is best determined from observations of HeII → HeI recombination lines in extragalactic HII (ionized hydrogen) regions. There is a good collection of abundance information on the $^4$He mass fraction, $Y$, O/H, and N/H in over 70 such regions[8,9]. Since $^4$He is produced in stars along with heavier elements such as Oxygen, it is then expected that the primordial abundance of $^4$He can be determined from the intercept of the correlation between $Y$ and O/H, namely $Y_p = Y(O/H \to 0)$. A detailed analysis[10] of the data found

$$Y_p = 0.238 \pm 0.002 \pm 0.005 \qquad (5)$$

The first uncertainty is purely statistical and the second uncertainty is an estimate of the systematic uncertainty in the primordial abundance determi-

Otai: submitted to **World Scientific** on October 27, 2018     7

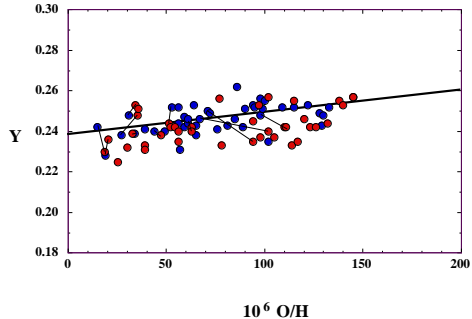

Figure 5. The Helium (Y) and Oxygen (O/H) abundances in extragalactic HII regions, from refs. [8] and [9]. Lines connect the same regions observed by different groups. The regression shown leads to the primordial $^4$He abundance given in Eq. (5).

nation. The solid box for $^4$He in Figure 4 represents the range (at $2\sigma_{\rm stat}$) from (5). The dashed box extends this by including the systematic uncertainty. The $^4$He data is shown in Figure 5. Here one sees the correlation of $^4$He with O/H and the linear regression which leads to primordial abundance given in Eq. (5).

The helium abundance used to derive (5) was determined using assumed electron densities $n$ in the HII regions obtained from SII data. Izotov, Thuan, & Lipovetsky[9] proposed a method based on several He emission lines to "self-consistently" determine the electron density. Their data using this method yields a higher primordial value

$$Y_p = 0.244 \pm 0.002 \pm 0.005 \qquad (6)$$

One should also note that a recent determination[11] of the $^4$He abundance in a single object (the SMC) also using the self consistent method gives a primordial abundance of $0.234 \pm 0.003$ (actually, they observe $Y = 0.240 \pm 0.002$ at [O/H] = -0.8, where the abundance [O/H] is defined as the log of the abundance relative to the solar abundance, $[X/H] \equiv \log([X/H]/[X/H]_\odot)$). Indeed, this work was extended in a reanalysis[12] of selected regions from ref. [9] and argue for corrections which lower the abundance to $0.239 \pm 0.002$ based on a fit to 5 regions.

As one can see, the resulting primordial $^4$He abundance shows significant sensitivity to the *method* of abundance determination, leading one to conclude that the systematic uncertainty (which is already dominant) may be underestimated[13].

## 2.3 Data-$^7$Li

The abundance of $^7$Li has been determined by observations of over 100 hot, population-II stars, and is found to have a very nearly uniform abundance[14]. For stars with a surface temperature $T > 5500$ K and a metallicity less than about 1/20th solar (so that effects such as stellar convection may not be important), the abundances show little or no dispersion beyond that which is consistent with the errors of individual measurements. There is, however, an important source of systematic error due to the possibility that Li has been depleted in these stars, though the lack of dispersion in the Li data limits the amount of depletion. In fact, a small observed slope in Li vs Fe and the tiny dispersion about that correlation indicates that depletion is negligible in these stars [15]. Furthermore, the slope may indicate a lower abundance of Li than that in (6). The observation[16] of the fragile isotope $^6$Li is another good indication that $^7$Li has not been destroyed in these stars[17].

The weighted mean of the $^7$Li abundance in the sample of ref. [15] is [Li] = 2.12 ([Li] = log $^7$Li/H + 12). It is common to test for the presence of a slope in the Li data by fitting a regression of the form [Li] = $\alpha + \beta$ [Fe/H]. These data indicate a rather large slope, $\beta = 0.07 - 0.16$ and hence a downward shift in the "primordial" lithium abundance $\Delta$[Li] = $-0.20 - -0.09$. Models of galactic evolution which predict a small slope for [Li] vs. [Fe/H], can produce a value for $\beta$ in the range $0.04 - 0.07$ [18]. Overall, when the regression based on the data and other systematic effects are taken into account a best value for Li/H was found to be[18]

$$\text{Li/H} = 1.23 \pm 0.1 \times 10^{-10} \tag{7}$$

with a plausible range between $0.9 - 1.9 \times 10^{-10}$. The dashed box in Figure 4 corresponds to this range in Li/H.

Figure 6 shows the different Li components for a model with $(^7\text{Li/H})_p = 1.23 \times 10^{-10}$. The linear slope produced by the model is independent of the input primordial value (unlike the log slope given above). The model of ref. [19] includes in addition to primordial $^7$Li, lithium produced in galactic cosmic ray nucleosynthesis (primarily $\alpha + \alpha$ fusion), and $^7$Li produced by the $\nu$-process during type II supernovae. As one can see, these processes are not sufficient to reproduce the population I abundance of $^7$Li, and additional production sources are needed.

## 2.4 Likelihood Analyses

At this point, having established the primordial abundance of at least two of the light elements, $^4$He and $^7$Li, with reasonable certainty, it is possible to test the concordance of BBN theory with observations. Two elements are sufficient for not only constraining the one parameter theory of BBN, but also for testing for consistency[20]. A theoretical likelihood function for $^4$He can be defined as

$$L_{\rm BBN}(Y, Y_{\rm BBN}) = e^{-(Y-Y_{\rm BBN}(\eta))^2/2\sigma_1^2} \qquad (8)$$

where $Y_{\rm BBN}(\eta)$ is the central value for the $^4$He mass fraction produced in the big bang as predicted by the theory at a given value of $\eta$. $\sigma_1$ is the uncertainty in that value derived from the Monte Carlo calculations [3,7] and is a measure of the theoretical uncertainty in the BBN calculation. Similarly one can write down an expression for the observational likelihood function. Assuming Gaussian errors, the likelihood function for the observations would take a form similar to that in (8).

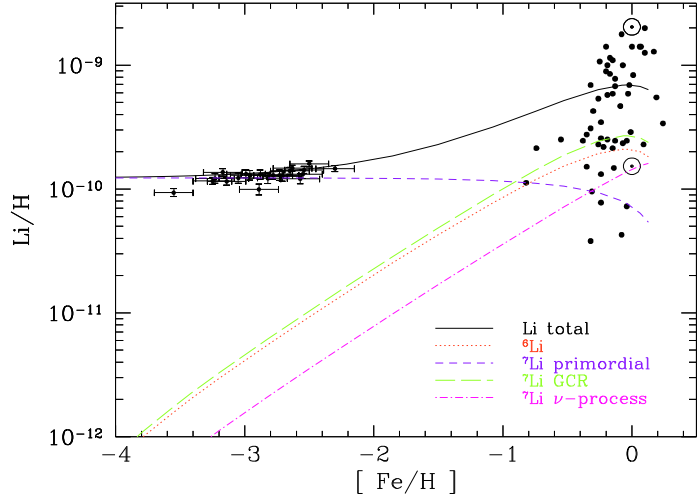

Figure 6. Contributions to the total predicted lithium abundance from the adopted GCE model of ref. [19], compared with low metallicity stars (from ref. [15]) and a sample of high metallicity stars. The solid curve is the sum of all components.

A total likelihood function for each value of $\eta$ is derived by convolving the theoretical and observational distributions, which for $^4$He is given by

$$L^{^4\text{He}}{}_{\text{total}}(\eta) = \int dY L_{\text{BBN}}\left(Y, Y_{\text{BBN}}(\eta)\right) L_{\text{O}}(Y, Y_{\text{O}}) \tag{9}$$

An analogous calculation is performed[20] for $^7$Li. The resulting likelihood functions from the observed abundances given in Eqs. (5) and (7) is shown in Figure 7. As one can see there is very good agreement between $^4$He and $^7$Li in the range of $\eta_{10} \simeq 1.5$ – $5.0$.

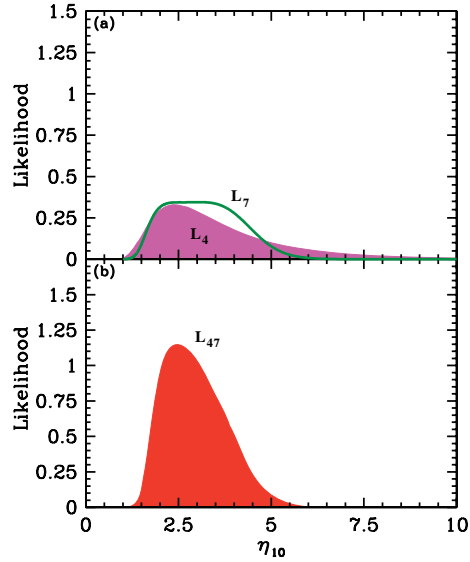

Figure 7. Likelihood distribution for each of $^4$He and $^7$Li, shown as a function of $\eta$ in the upper panel. The lower panel shows the combined likelihood function.

The combined likelihood, for fitting both elements simultaneously, is given by the product of the two functions in the upper panel of Figure 7 and is shown in the lower panel. The 95% CL region covers the range $1.7 < \eta_{10} < 4.7$, with the peak value occurring at $\eta_{10} = 2.4$. This range corresponds to values of $\Omega_B h^2$ between

$$0.006 < \Omega_B h^2 < 0.017 \tag{10}$$

with a central value of $\Omega_B h^2 = 0.009$. Using the higher value for $^4$He ($Y_p =$ 0.244) would result in an upward shift in $\eta_{10}$ by about 0.1 – 0.2.

## 2.5 Data-D/H

The remaining two light elements produced by BBN are D and $^3$He. For the most part, the abundances of these elements are determined either in the local interstellar medium or in our own solar system. As such, one needs a model of galactic chemical evolution to tie them to the BBN abundances. Deuterium is predicted to be a monotonically decreasing function of time. The degree to which D is destroyed, however, is a model dependent question and related to the production of $^3$He. Stellar models predict that substantial amounts of $^3$He are produced in stars between 1 and 3 M$_\odot$[21]. It should be emphasized that this prediction is in fact consistent with the observation of high $^3$He/H in planetary nebulae[22]. However, the implementation of standard model $^3$He yields in chemical evolution models leads to an overproduction of $^3$He/H particularly at the solar epoch[23]. See ref. [24] for attempts to resolve this problem. Because of this model dependence, I will not consider $^3$He further here.

Despite the problem of relating many of the locally observed D/H measurements to BBN, we can use the ISM values to set a firm lower limit on primordial D/H, and hence an upper limit to $\eta$ because of the monotonically decreasing history of D/H in the Galaxy. This is shown as the dashed (half)-box in Fig. 4.

There have been several reported measurements of D/H in high redshift quasar absorption systems. Such measurements are in principle capable of determining the primordial value for D/H and hence $\eta$, because of the strong and monotonic dependence of D/H on $\eta$. However, at present, detections of D/H using quasar absorption systems do not yield a conclusive value for D/H. In addition to the earlier determinations[25] in two Lyman limit systems of $4.0 \pm .65$ and $3.25 \pm 0.3 \times 10^{-5}$ and the upper limit in a third Lyman limit system[26] of $6.8 \times 10^{-5}$, there have been three new measurements of D/H in Damped Lyman $\alpha$ systems: $2.5 \pm 0.25$ [27]; $2.25 \pm 0.65$ [28]; and $1.65 \pm 0.35$ [29]; all times $10^{-5}$. This data is shown in Fig. 8. Also shown in Fig. 8, are the adjusted D/H abundances when more complex velocity distributions are included as in refs. [30].

Is there a real dispersion in D/H in these high redshift systems? There are two possible trends that go in a similar direction. The data may show an inverse correlation of D/H abundance with Si [27,29]. This may be an artifact of poorly determined Si abundances, or (as yet unknown) systematics affecting the D/H determination in high-column density (damped Lyman-$\alpha$, hereafter DLA) or low-column density (Lyman limit systems) absorbers. On the other hand, if the correlation is real it would indicate that chemical evolution processes have occurred in these systems. The second trend is that the data may show an inverse correlation of D/H abundance with HI column density. If real, this would suggest that in the high column density DLA systems, which are most likely to have undergone some star formation, some processing of D/H must similarly have occurred at high redshift. One can only conclude: if the dispersion in D/H is real, it has profound consequences, as it indicates that *some processing of D/H must have occurred even at high redshift.*

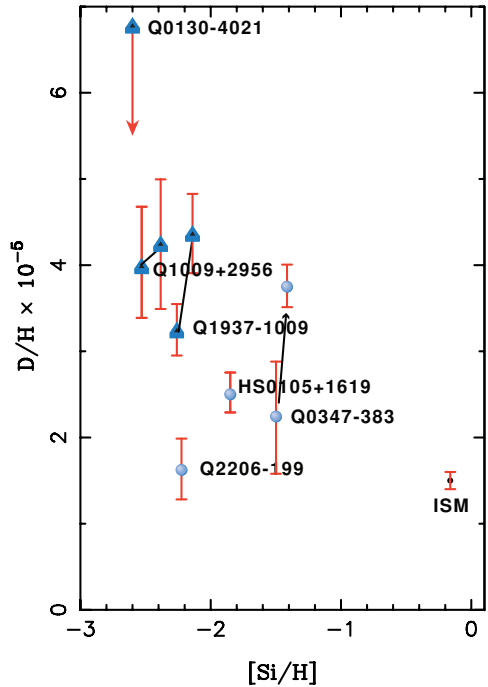

Figure 8. The D/H data as a function of metallicity given by [Si/H].

It is interesting to speculate[31] that the possible high redshift destruction of D/H is related to recent observations which suggest the existence of a white dwarf population in the Galactic halo[32]. These observations could be signatures of an early population of intermediate-mass stars. Such a population requires a Population III initial mass function different from that of the solar neighborhood. Also, to avoid overproduction of C and N, it is required that the $Z = 0$ yields of these stars have low ($\sim 10^{-3}$ solar) abundances as suggested by some recent calculations. Under these assumptions, it is possible to model the observed D vs Si trend[31]. Such a scenario predicts a high cosmic Type Ia supernova rate, while producing a white dwarf population that accounts for only $\sim 1.5\%$ of the dark halo.

It is clear that a simple average of D/H abundance determinations does not make sense, at least without a proper enlargement of the error in the mean due to the poor $\chi^2$ that such a mean would produce. Moreover, if deuterium destruction has occurred, we must question the extent to which any of these systems determine the value of $\Omega_B$. It is important to note however, that for the upper end of the range ($\sim 5 \times 10^{-5}$) shown in Fig. 8, all of the element abundances are consistent as will be discussed below.

## 2.6 More Analysis

It is interesting to compare the results from the likelihood functions of $^4$He and $^7$Li with that of D/H. To include D/H, one would proceed in much the same way as with the other two light elements. We compute likelihood functions for the BBN predictions as in Eq. (8) and the likelihood function for the observations. These are then convolved as in Eq. (9).

Using D/H $= (3.0 \pm 0.3) \times 10^{-5}$ as the primordial abundance, one obtains the likelihood function shown (shaded) in the upper panel of Fig. 9. The 95% CL region covers the range $5.0 < \eta_{10} < 7.4$, with the peak value occurring at $\eta_{10} = 5.8$. This range corresponds to values of $\Omega_B$ between

$$0.018 < \Omega_B h^2 < 0.027 \qquad (11)$$

with a central value of $\Omega_B h^2 = 0.021$.

The combined likelihood, for fitting all elements (D, $^4$He, and $^7$Li) simultaneously, is shown by the shaded curve in the lower panel of Figure 9. Note that it has been scaled upwards by a factor of 25. In this case, the 95% CL region covers the range $4.7 < \eta_{10} < 6.2$, with the peak value occurring at $\eta_{10} = 5.3$. This range corresponds to values of $\Omega_B$ between

$$0.017 < \Omega_B h^2 < 0.023 \qquad (12)$$

with a central value of $\Omega_B h^2 = 0.019$.

It is important to recall however, that the true uncertainty in the low D/H systems might be somewhat larger. If we allow D/H to be as large as $5 \times 10^{-5}$, the peak of the D/H likelihood function shifts down to $\eta_{10} \simeq 4$. In this case, there would be a near perfect overlap with the high $\eta$ $^7$Li peak and since the $^4$He distribution function is very broad, this would be a highly compatible solution. Given our discussion in the previous section concerning the current status of the D/H data, it is premature to claim a lack of concordance between BBN theory and observations.

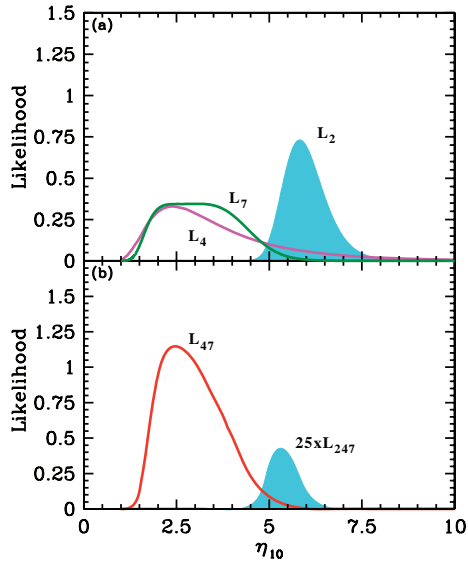

Figure 9. Likelihood distribution for D/H, shown (shaded) as a function of $\eta$ in the upper panel. Also shown are the $^4$He and $^7$Li likelihoods from Fig. 7. The lower panel shows the combined likelihood function (shaded) compared to the previous case neglecting D/H.

## 3   The CMB - BBN connection

It is interesting to note the role of BBN in the prediction of the microwave background[1]. The argument is rather simple. BBN requires temperatures greater than 100 keV, which according to the standard model time-temperature relation, $t_s T_{\rm MeV}^2 = 2.4/\sqrt{N}$, where $N$ is the number of relativistic degrees of freedom at temperature $T$, and corresponds to timescales less than about 200 s. The typical cross section for the first link in the nucleosynthetic

chain is

$$\sigma v(p + n \to D + \gamma) \simeq 5 \times 10^{-20} \text{cm}^3/\text{s} \qquad (13)$$

This implies that it was necessary to achieve a density

$$n \sim \frac{1}{\sigma v t} \sim 10^{17} \text{cm}^{-3} \qquad (14)$$

The density in baryons today is known approximately from the density of visible matter to be $n_{Bo} \sim 10^{-7}$ cm$^{-3}$ and since we know that that the density $n$ scales as $R^{-3} \sim T^3$, the temperature today must be

$$T_o = (n_{Bo}/n)^{1/3} T_{\text{BBN}} \sim 10\text{K} \qquad (15)$$

thus linking two of the most important tests of the Big Bang theory.

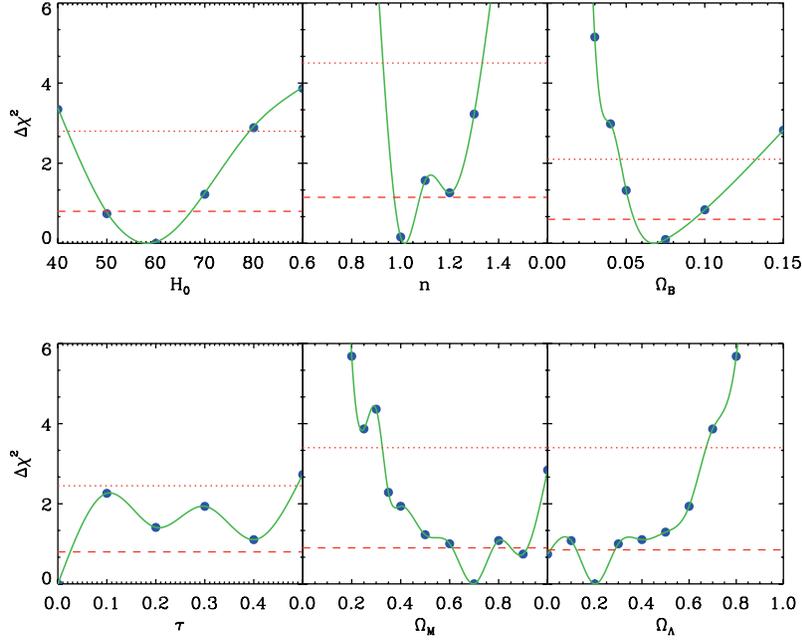

Figure 10. $\Delta\chi^2$ calculated with the MAXIMA-1 and COBE data as a function of parameter value. Solid blue circles show grid points in parameter space, and the green lines were obtained by interpolating between grid points. The parameter values where the green line intercepts the red dashed (dotted) line corresponds to the 68% (95%) frequentist confidence region[39].

Microwave background anisotropy measurements have made tremendous advances in the last few years. The power spectrum[33,34,35] has been measured relatively accurately out to multipole moments corresponding to $\ell \sim 1000$. The details of this spectrum enable one to make accurate predictions of a large number of fundamental cosmological parameters[34,36,37,38]. An example of these results as found by a recent frequentist analysis[39] is shown in Fig. 10.

The CMB anisotropies thus *independently test* the BBN prediction of $\Omega_B h^2$. At present, the predicted BBN baryon densities from D/H agree to an uncanny level with the most recent CMB results[34,37]. The recent result from DASI[37] indicates that $\Omega_B h^2 = 0.022^{+0.004}_{-0.003}$, while that of BOOMERanG-98 [34], $\Omega_B h^2 = 0.021^{+0.004}_{-0.003}$ (using $1\sigma$ errors) which should be compared to the BBN prediction given in eq. 11. These determinations are lower than value found by MAXIMA-1 [36] which yields $\Omega_B h^2 = 0.0325 \pm 0.006$. Given the current uncertainties, these results are consistent as can be seen in Fig. 11 based on the recent frequentist analysis[39] which found $\Omega_B h^2 = 0.026^{+0.010}_{-0.006}$ In addition, the BOOMERanG and DASI determinations are higher than the value $\Omega_B h^2 = 0.009$ based on $^4$He and $^7$Li. However, the measurements of the Cosmic Background Imager[38] at smaller angular scales (higher multipoles) agree with lower BBN predictions and claims a maximum likelihood value for $\Omega_B h^2 = 0.009$ (albeit with a large uncertainty).

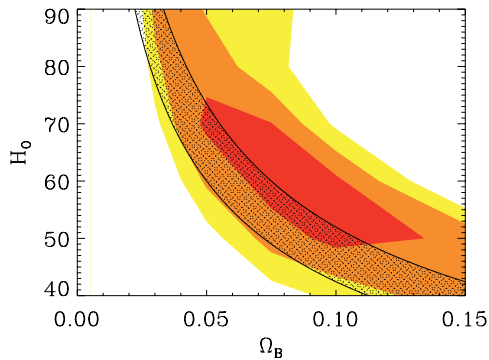

Figure 11. Two-dimensional frequentist confidence regions in the $(H_0, \Omega_B)$ plane[39]. The red, orange and yellow regions correspond to the 68%, 95%, and 99% confidence regions respectively. Standard calculations from big bang nucleosynthesis and observations of $\frac{D}{H}$ predict a 95% confidence region of $\Omega_B h^2 = 0.021^{+0.006}_{-0.003}$, indicated by the shaded region.

## 4 Constraints from BBN

Limits on particle physics beyond the standard model are mostly sensitive to the bounds imposed on the $^4$He abundance. As discussed earlier, the neutron-to-proton ratio is fixed by its equilibrium value at the freeze-out of the weak interaction rates at a temperature $T_f \sim 1$ MeV modulo the occasional free neutron decay. Furthermore, freeze-out is determined by the competition between the weak interaction rates and the expansion rate of the Universe

$$G_F^2 T_f^5 \sim \Gamma_{\text{wk}}(T_f) = H(T_f) \sim \sqrt{G_N N} T_f^2 \qquad (16)$$

where $N$ counts the total (equivalent) number of relativistic particle species. At $T \sim 1$ MeV, $N = 43/4$. The presence of additional neutrino flavors (or any other relativistic species) at the time of nucleosynthesis increases the overall energy density of the Universe and hence the expansion rate leading to a larger value of $T_f$, $(n/p)$, and ultimately $Y_p$. Because of the form of Eq. (16) it is clear that just as one can place limits[40] on $N$, any changes in the weak or gravitational coupling constants can be similarly constrained (for a discussion see ref. [41]).

Changes in $N_\nu$ actually affect not only $^4$He, but also the abundances of the other light elements as seen in Fig. 12 [42]. We see the typical large dependence on $N_\nu$ in $^4$He, but also note the shifts in the other elements, particularly D, and also Li over some ranges in $\eta$. Because of these variations, one is not restricted to only $^4$He in testing $N_\nu$ and particle physics.

Just as $^4$He and $^7$Li were sufficient to determine a value for $\eta$, a limit on $N_\nu$ can be obtained as well[20,43,44]. The likelihood approach utilized above can be extended to include $N_\nu$ as a free parameter. Since the light element abundances can be computed as functions of both $\eta$ and $N_\nu$, the likelihood function can be defined by[43] replacing the quantity $Y_{\text{BBN}}(\eta)$ in eq. (8) with $Y_{\text{BBN}}(\eta, N_\nu)$ to obtain $L^{^4\text{He}}{}_{\text{total}}(\eta, N_\nu)$. Again, similar expressions are needed for $^7$Li and D.

The likelihood distribution derived from the analysis of ref. [42] is shown in Fig. 13 where iso-likelihood contours representing 68, 95, and 99 % CL are projected onto the $\eta_{10} - N_\nu$ plane. As one can see, when only $^4$He and $^7$Li are used, the allowed range in $\eta_{10}$ is rather broad and the upper limit to $N_\nu \lesssim 4.2$ at relatively low $\eta$. At present, one can not use D/H to fix $N_\nu$, unless one has additional information on the value of $\eta$, e.g., from the CMB as is demonstrated in Fig. 14. On the other hand when all three light elements are used, the narrow range in $\eta$ at relatively high values, allows for a tight constraint on $N_\nu <$ at 95 % CL. Furthermore, it is clear that the most realistic value, $N_\nu = 3$ is in fact consistent with all of the observed abundances at the

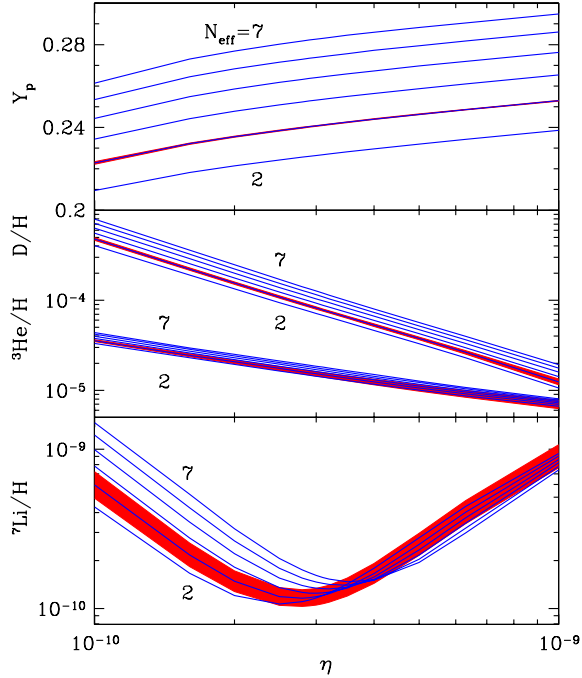

Figure 12. BBN abundance predictions[42] as a function of the baryon-to-photon ratio $\eta$, for $N_\nu = 2$ to 7. The bands show the $1\sigma$ error bars. Note that for the isotopes other than Li, the error bands are comparable in width to the thickness of the abundance curve shown. All bands are centered on $N_\nu = 3$.

$2\ \sigma$ level.

Of course as discussed above, the CMB provides an independent determination of $\eta$, we can use that to fix $\eta$, and test the available constraints on $N_\nu$ [42]. For example, we can fix $\eta = 5.8 \times 10^{-10}$ which corresponds to the BOOMERanG and DASI determinations, and use the current $^4$He and $^7$Li data to compute a BBN likelihood function for $N_\nu$. This is shown in the first panel of Fig. 14. Assuming a 10% uncertainty in $\eta$, the 95 % CL range on $N_\nu$ is 1.8 – 3.3, whereas the 1 sided upper limit with a $N_\nu \geq 3$ prior[45] is 3.5. Using instead the D/H data, at the same baryon density, one finds an upper limit $N_\nu < 6.3$. This likelihood is shown in panel b) of Fig. 14. Alternatively, we could choose a lower value of $\eta = 2.4 \times 10^{-5}$, and using $^4$He and $^7$Li we find that the 95 % CL range is 2.2 – 3.9. This case is exemplified in panel c).

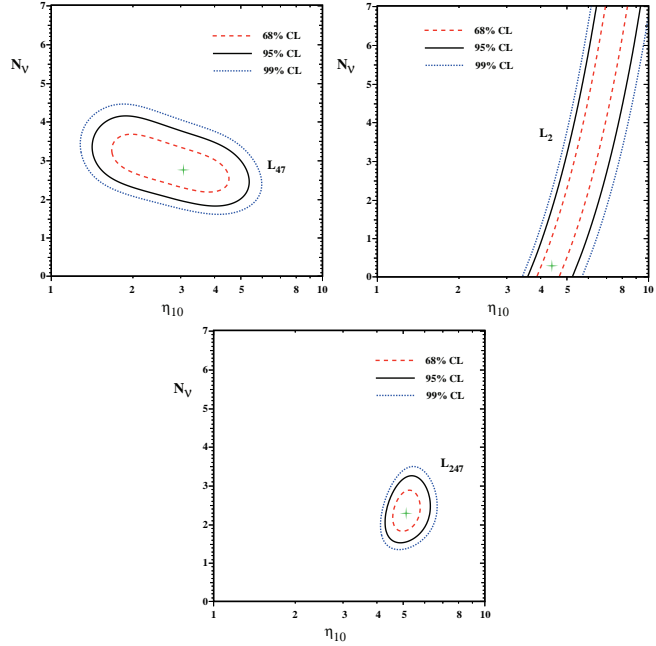

Figure 13. (a) Likelihood contours representing 68, 95, and 99 % CL projected onto the $\eta_{10} - N_\nu$ plane. In a) only $^4$He and $^7$Li are used. In b) only D is used, and in c) all three light elements are used. The crosses denote the position of the peak of the likelihood distribution.

## 5 Something's Missing

There is considerable evidence for dark matter in the Universe[46]. The best observational evidence is found on the scale of galactic halos and comes from the observed flat rotation curves of galaxies. There is also good evidence for dark matter in elliptical galaxies, as well as clusters of galaxies coming from X-ray observations of these objects. In theory, we expect dark matter because 1) inflation predicts $\Omega = 1$, and the upper limit on the baryon (visible) density of the Universe from big bang nucleosynthesis is $\Omega_B < 0.1$; 2) Even in the absence of inflation (which does not distinguish between matter and a cosmological constant), the large scale power spectrum is consistent with a cosmological matter density of $\Omega \sim 0.3$, still far above the limit from nucleosynthesis; and 3) our current understanding of galaxy formation is inconsistent with observations if the Universe is dominated by baryons.

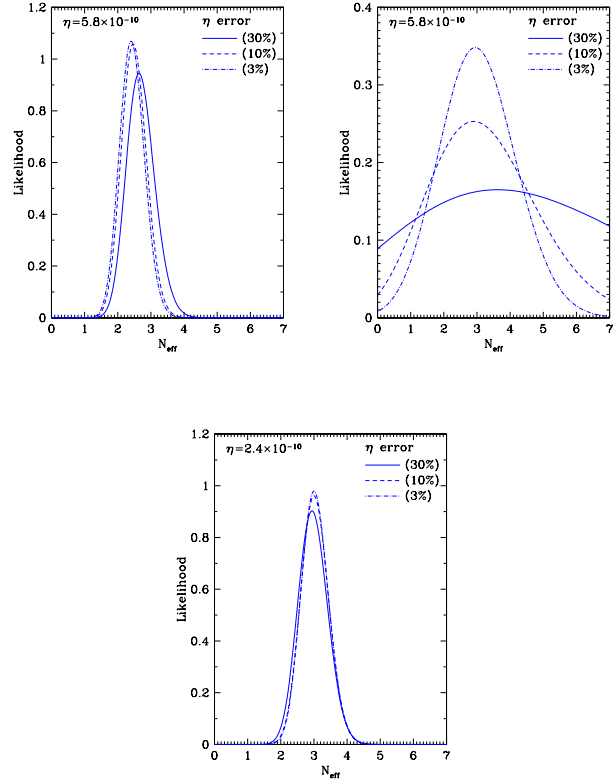

Figure 14. (a) The distribution in $N_\nu$ assuming a value of $\eta = 5.8 \times 10^{-10}$ and primordial $^4$He & $^7$Li abundances as in eqs. 5 and 7. (b) As in (a) assuming a D measurement at the current precision (taken to be $3.0 \pm 0.3 \times 10^{-5}$). (c) As in (a), but with a value for $\eta = 2.4 \times 10^{-10}$ .

Indeed, we now have direct evidence from CMB anisotropy measurements that $\Omega_{\text{tot}} = 1$ to with in about 20 %. The CMB anisotropy allows for a determination of the curvature of the Universe. Therefore, while one can not determine unambiguously the value of $\Omega_{\text{matter}}$, one can fix the sum of the matter and dark energy (cosmological constant) contributions. This is shown in Fig. 15 from the frequentist analysis in[39]. When combined with high redshift supernovae measurements indicating that the Universe is currently accelerating[47], one would conclude that the dark energy constitutes about 65 % of the closure density leaving the remaining 35 % for matter. Given that the baryon density is less than 5 %, this leaves us with about 30 % of closure density for non-baryonic dark matter. Other dynamical arguments estimating the density of matter are in agreement with this allotment (see ref. [48] for a recent review).

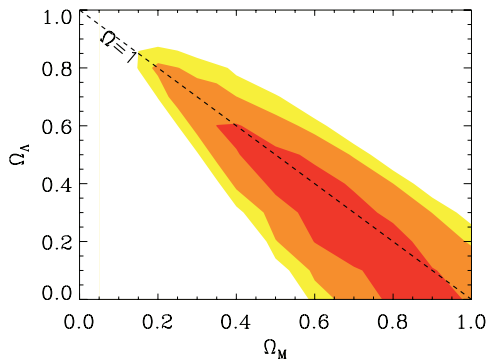

Figure 15. Two-dimensional frequentist confidence regions in the $(\Omega_M, \Omega_\Lambda)$ plane. The red, orange and yellow regions correspond to the 68%, 95%, and 99% confidence regions respectively. The dashed black line corresponds to a flat universe, $\Omega = \Omega_M + \Omega_\Lambda = 1$.

In fact, there are many reasons why most of the dark matter must be non-baryonic. In addition to the problems with baryonic dark matter associated with nucleosynthesis or the growth of density perturbations, it is very difficult to hide baryons. There are indeed very good constraints on the possible forms for baryonic dark matter in our galaxy. Strong cases can be made against hot gas, dust, jupiter size objects, and stellar remnants such as white dwarfs and neutron stars[49].

In what follows, I will focus on supersymmetric candidates in which the relic abundance of dark matter contributes a significant though not excessive amount to the overall energy density. Denoting by $\Omega_\chi$ the fraction of the critical energy density provided by the dark matter, the density of interest

falls in the range

$$0.1 \leq \Omega_\chi h^2 \leq 0.3 \tag{17}$$

The lower limit in eq.(17) is motivated by astrophysical relevance. For lower values of $\Omega_\chi h^2$, there is not enough dark matter to play a significant role in structure formation, or constitute a large fraction of the critical density. The upper bound in (17), on the other hand, is an absolute constraint, derivable from the age of the Universe, which can be expressed as

$$H_0 t_0 = \int_0^1 dx \left(1 - \Omega - \Omega_\Lambda + \Omega_\Lambda x^2 + \Omega/x\right)^{-1/2} \tag{18}$$

Given a lower bound on the age of the Universe, one can establish an upper bound on $\Omega h^2$ from eq.(18). The limit $t_0 \gtrsim 12$ Gyr translates into the upper bound given in (17). Adding a cosmological constant does not relax the upper bound on $\Omega h^2$, so long as $\Omega + \Omega_\Lambda \leq 1$. If indeed, the indications for a cosmological constant from recent supernovae observations[47] turn out to be correct, the density of dark matter will be constrained to the lower end of the range in (17). Indeed for $h^2 \sim 1/2$, we expect $\Omega_\chi h^2 \simeq 0.15$.

## 6 Supersymmetric Dark Matter

Although there are many reasons for considering supersymmetry as a candidate extension to the standard model of strong, weak and electromagnetic interactions[50], one of the most compelling is its role in understanding the hierarchy problem[51] namely, why/how is $m_W \ll M_P$. One might think naively that it would be sufficient to set $m_W \ll M_P$ by hand. However, radiative corrections tend to destroy this hierarchy. For example, one-loop diagrams generate

$$\delta m_W^2 = \mathcal{O}\left(\frac{\alpha}{\pi}\right) \Lambda^2 \gg m_W^2 \tag{19}$$

where $\Lambda$ is a cut-off representing the appearance of new physics, and the inequality in (19) applies if $\Lambda \sim 10^3$ TeV, and even more so if $\Lambda \sim m_{GUT} \sim 10^{16}$ GeV or $\sim M_P \sim 10^{19}$ GeV. If the radiative corrections to a physical quantity are much larger than its measured values, obtaining the latter requires strong cancellations, which in general require fine tuning of the bare input parameters. However, the necessary cancellations are natural in supersymmetry, where one has equal numbers of bosons $B$ and fermions $F$ with equal couplings, so that (19) is replaced by

$$\delta m_W^2 = \mathcal{O}\left(\frac{\alpha}{\pi}\right) |m_B^2 - m_F^2| . \tag{20}$$

The residual radiative correction is naturally small if $|m_B^2 - m_F^2| \lesssim 1$ TeV$^2$.

In order to justify the absence of superpotential terms which can be responsible for extremely rapid proton decay, it is common in the minimal supersymmetric standard model (MSSM) to assume the conservation of R-parity. If R-parity, which distinguishes between "normal" matter and the supersymmetric partners and can be defined in terms of baryon, lepton and spin as $R = (-1)^{3B+L+2S}$, is unbroken, there is at least one supersymmetric particle (the lightest supersymmetric particle or LSP) which must be stable.

The stability of the LSP almost certainly renders it a neutral weakly interacting particle[52]. Strong and electromagnetically interacting LSPs would become bound with normal matter forming anomalously heavy isotopes. Indeed, there are very strong upper limits on the abundances, relative to hydrogen, of nuclear isotopes[53], $n/n_H \lesssim 10^{-15}$ to $10^{-29}$ for 1 GeV $\lesssim m \lesssim$ 1 TeV. A strongly interacting stable relics is expected to have an abundance $n/n_H \lesssim 10^{-10}$ with a higher abundance for charged particles.

There are relatively few supersymmetric candidates which are not colored and are electrically neutral. The sneutrino[54] is one possibility, but in the MSSM, it has been excluded as a dark matter candidate by direct[55] and indirect[56] searches. In fact, one can set an accelerator based limit on the sneutrino mass from neutrino counting, $m_{\tilde{\nu}} \gtrsim 43$ GeV [57]. In this case, the direct relic searches in underground low-background experiments require $m_{\tilde{\nu}} \gtrsim 1$ TeV [58]. Another possibility is the gravitino which is probably the most difficult to exclude. I will concentrate on the remaining possibility in the MSSM, namely the neutralinos.

### 6.1 Relic Densities

There are four neutralinos, each of which is a linear combination of the $R = -1$ neutral fermions,[52]: the wino $\tilde{W}^3$, the partner of the 3rd component of the $SU(2)_L$ gauge boson; the bino, $\tilde{B}$, the partner of the $U(1)_Y$ gauge boson; and the two neutral Higgsinos, $\tilde{H}_1$ and $\tilde{H}_2$. Assuming gaugino mass universality at the GUT scale, the identity and mass of the LSP are determined by the gaugino mass $m_{1/2}$ (or equivalently by the SU(2) gaugino mass $M_2$ at the weak scale – at the GUT scale, $M_2 = m_{1/2}$), the Higgs mixing mass $\mu$, and the ratio of Higgs vevs, $\tan\beta$. In general, neutralinos can be expressed as a linear combination

$$\chi = \alpha \tilde{B} + \beta \tilde{W}^3 + \gamma \tilde{H}_1 + \delta \tilde{H}_2 \tag{21}$$

The solution for the coefficients $\alpha, \beta, \gamma$ and $\delta$ for neutralinos that make up the LSP can be found by diagonalizing the mass matrix

$$(\tilde{W}^3, \tilde{B}, \tilde{H}_1^0, \tilde{H}_2^0) \begin{pmatrix} M_2 & 0 & \frac{-g_2 v_1}{\sqrt{2}} & \frac{g_2 v_2}{\sqrt{2}} \\ 0 & M_1 & \frac{g_1 v_1}{\sqrt{2}} & \frac{-g_1 v_2}{\sqrt{2}} \\ \frac{-g_2 v_1}{\sqrt{2}} & \frac{g_1 v_1}{\sqrt{2}} & 0 & -\mu \\ \frac{g_2 v_2}{\sqrt{2}} & \frac{-g_1 v_2}{\sqrt{2}} & -\mu & 0 \end{pmatrix} \begin{pmatrix} \tilde{W}^3 \\ \tilde{B} \\ \tilde{H}_1^0 \\ \tilde{H}_2^0 \end{pmatrix} \quad (22)$$

where $M_1$ is a soft supersymmetry breaking term giving mass to the U(1) gaugino. In a unified theory $M_1 = M_2$ at the unification scale (at the weak scale, $M_1 = \frac{5}{3}\frac{\alpha_1}{\alpha_2}M_2$). As one can see, the coefficients $\alpha, \beta, \gamma$, and $\delta$ depend only on $m_{1/2}$, $\mu$, and $\tan\beta$.

In Figure 16 [59], regions in the $M_2, \mu$ plane with $\tan\beta = 2$ are shown in which the LSP is one of several nearly pure states, the photino, $\tilde{\gamma}$, the bino, $\tilde{B}$, a symmetric combination of the Higgsinos, $\tilde{H}_{(12)}$, or the Higgsino, $\tilde{S} = \sin\beta \tilde{H}_1 + \cos\beta \tilde{H}_2$. The dashed lines show the LSP mass contours. The cross hatched regions correspond to parameters giving a chargino $(\tilde{W}^\pm, \tilde{H}^\pm)$ state with mass $m_{\tilde{\chi}} \leq 45 GeV$ and as such are excluded by LEP[60]. This constraint has been extended by LEP[61] and is shown by the light shaded region and corresponds to regions where the chargino mass is $\lesssim 104$ GeV. The newer limit does not extend deep into the Higgsino region because of the degeneracy between the chargino and neutralino. Notice that the parameter space is dominated by the $\tilde{B}$ or $\tilde{H}_{12}$ pure states and that the photino only occupies a small fraction of the parameter space, as does the Higgsino combination $\tilde{S}$. Both of these light states are now experimentally excluded.

The relic abundance of LSP's is determined by solving the Boltzmann equation for the LSP number density in an expanding Universe. The technique[62] used is similar to that for computing the relic abundance of massive neutrinos[63]. The relic density depends on additional parameters in the MSSM beyond $M_2, \mu$, and $\tan\beta$. These include the sfermion masses, $m_{\tilde{f}}$, the Higgs pseudo-scalar mass, $m_A$, and the tri-linear masses $A$ as well as two phases $\theta_\mu$ and $\theta_A$. To determine, the relic density it is necessary to obtain the general annihilation cross-section for neutralinos. In much of the parameter space of interest, the LSP is a bino and the annihilation proceeds mainly through sfermion exchange as shown in Figure 17. For binos, it is possible to adjust the sfermion masses to obtain closure density in a wide mass range. Adjusting the sfermion mixing parameters[64] or CP violating phases[65,66] allows even greater freedom.

Because of the p-wave suppression associated with Majorana fermions, the s-wave part of the annihilation cross-section is suppressed by the outgoing fermion masses. This means that it is necessary to expand the cross-section to include p-wave corrections which can be expressed as a term proportional to the temperature if neutralinos are in equilibrium. Unless the neutralino mass happens to lie near near a pole, such as $m_\chi \simeq m_Z/2$ or $m_h/2$, in which case there are large contributions to the annihilation through direct $s$-channel resonance exchange, the dominant contribution to the $\tilde{B}\tilde{B}$ annihilation cross section comes from crossed $t$-channel sfermion exchange. In the absence of such a resonance, the thermally-averaged cross section for $\tilde{B}\tilde{B} \to f\bar{f}$ takes the generic form

$$\langle \sigma v \rangle = (1 - \frac{m_f^2}{m_{\tilde{B}}^2})^{1/2} \frac{g_1^4}{128\pi} \left[ (Y_L^2 + Y_R^2)^2 (\frac{m_f^2}{\Delta_f^2}) \right.$$
$$\left. + (Y_L^4 + Y_R^4)(\frac{4m_{\tilde{B}}^2}{\Delta_f^2})(1 + ...)\, x \right]$$
$$\equiv a + bx \tag{23}$$

where $Y_{L(R)}$ is the hypercharge of $f_{L(R)}$, $\Delta_f \equiv m_{\tilde{f}}^2 + m_{\tilde{B}}^2 - m_f^2$, and we have shown only the leading $P$-wave contribution proportional to $x \equiv T/m_{\tilde{B}}$. Such an expansion yields very accurate results unless the LSP is near a threshold, near a resonance, or is nearly degenerate in mass with another SUSY particle in which cases a more accurate treatment is necessary[67]. Annihilations in the early Universe continue until the annihilation rate $\Gamma \simeq \sigma v n_\chi$ drops below the expansion rate, $H$. For particles which annihilate through approximate weak scale interactions, this occurs when $T \sim m_\chi/20$. Subsequently, the relic density of neutralinos is fixed relative to the number of relativistic particles.

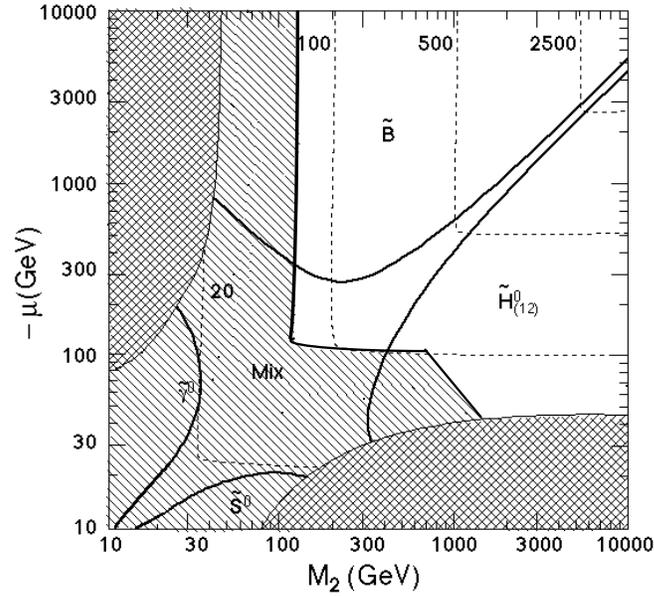

Figure 16. Mass contours and composition of nearly pure LSP states in the MSSM [59].

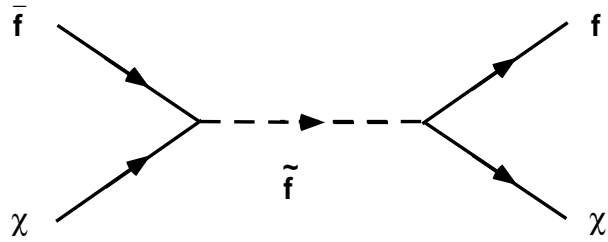

Figure 17. Typical annihilation diagram for neutralinos through sfermion exchange.

As noted above, the number density of neutralinos is tracked by a Boltzmann-like equation,

$$\frac{dn}{dt} = -3\frac{\dot{R}}{R}n - \langle \sigma v \rangle (n^2 - n_0^2) \tag{24}$$

where $n_0$ is the equilibrium number density of neutralinos. By defining the quantity $f = n/T^3$, we can rewrite this equation in terms of $x$, as

$$\frac{df}{dx} = m_\chi \left( \frac{1}{90}\pi^2 \kappa^2 N \right)^{1/2} (f^2 - f_0^2) \tag{25}$$

The solution to this equation at late times (small $x$) yields a constant value of $f$, so that $n \propto T^3$. The final relic density expressed as a fraction of the critical energy density can be written as[52]

$$\Omega_\chi h^2 \simeq 1.9 \times 10^{-11} \left(\frac{T_\chi}{T_\gamma}\right)^3 N_f^{1/2} \left(\frac{\text{GeV}}{ax_f + \frac{1}{2}bx_f^2}\right) \tag{26}$$

where $(T_\chi/T_\gamma)^3$ accounts for the subsequent reheating of the photon temperature with respect to $\chi$, due to the annihilations of particles with mass $m < x_f m_\chi$ [68]. The subscript $f$ refers to values at freeze-out, i.e., when annihilations cease.

In Figure 18 [69], regions in the $M_2 - \mu$ plane (rotated with respect to Figure 16) with $\tan \beta = 2$, and with a relic abundance $0.1 \leq \Omega h^2 \leq 0.3$ are shaded. In Figure 18, the sfermion masses have been fixed such that $m_0 = 100$ GeV (the dashed curves border the region when $m_0 = 1000$ GeV). Clearly the MSSM offers sufficient room to solve the dark matter problem. In the higgsino sector $\tilde{H}_{12}$, additional types of annihilation processes known as coannihilations [67,70,71] between $\tilde{H}_{(12)}$ and the next lightest neutralino ($\tilde{H}_{[12]}$) must be included. These tend to significantly lower the relic abundance in much of this sector and as one can see there is little room left for Higgsino dark matter[69].

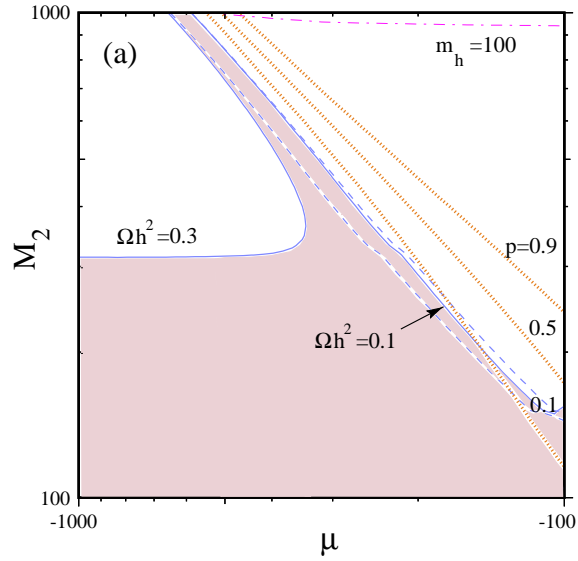

Figure 18. Regions in the $M_2$–$\mu$ plane where $0.1 \leq \Omega h^2 \leq 0.3$ [69]. Also shown are the Higgsino purity contours (labeled 0.1, 0.5, and 0.9). As one can see, the shaded region is mostly gaugino (low Higgsino purity). Masses are in GeV.

As should be clear from Figures 16 and 18, binos are a good and likely choice for dark matter in the MSSM. For fixed $m_{\tilde{f}}$, $\Omega h^2 \gtrsim 0.1$, for all $m_{\tilde{B}} = 20 - 250$ GeV largely independent of $\tan\beta$ and the sign of $\mu$. In addition, the requirement that $m_{\tilde{f}} > m_{\tilde{B}}$ translates into an upper bound of about 250 GeV on the bino mass[59,72]. By further adjusting the trilinear $A$ and accounting for sfermion mixing this upper bound can be relaxed[64] and by allowing for non-zero phases in the MSSM, the upper limit can be extended to about 600 GeV [65]. For fixed $\Omega h^2 = 1/4$, we would require sfermion masses of order 120 – 250 GeV for binos with masses in the range 20 – 250 GeV. The Higgsino relic density, on the other hand, is largely independent of $m_{\tilde{f}}$. For large $\mu$, annihilations into $W$ and $Z$ pairs dominate, while for lower $\mu$, it is the annihilations via Higgs scalars which dominate. Aside from a narrow region with $m_{\tilde{H}_{12}} < m_W$ and very massive Higgsinos with $m_{\tilde{H}_{12}} \gtrsim 500$ GeV, the relic density of $\tilde{H}_{12}$ is very low. Above about 1 TeV, these Higgsinos are also excluded.

As discussed in section 4, one can make a further reduction in the number of parameters by setting all of the soft scalar masses equal at the GUT scale (similarly for the $A$ parameters as well). In this case the free parameters are

$$m_0, m_{1/2}, A \quad \text{and} \quad \tan\beta\,, \tag{27}$$

with $\mu$ and $m_A$ being determined by the electroweak vacuum conditions, the former up to a sign. We refer to this scenario as the constrained MSSM (CMSSM). In Figure 19 [73], this parameter space is shown for $\tan\beta = 10$. The light shaded region corresponds to the portion of parameter space where the relic density $\Omega_\chi h^2$ is between 0.1 and 0.3. The dark shaded region (in the lower right) corresponds to the parameters where the LSP is not a neutralino but rather a $\tilde{\tau}_R$. The cosmologically interesting region at the left of the figure is due to the appearance of pole effects. There, the LSP can annihilate through s-channel $Z$ and $h$ (the light Higgs) exchange, thereby allowing a very large value of $m_0$. However, this region is excluded by the phenomenological constraints described below. The cosmological region extends toward large values of $m_{1/2}$, due to coannihilation effects between the lightest neutralino and the $\tilde{\tau}_R$ [78]. For non-zero values of $A_0$, there are new regions for which $\chi - \tilde{t}$ are important[79].

Figure 19 also shows the current experimental constraints on the CMSSM parameter space provided by direct searches at LEP. One of these is the limit $m_{\chi^\pm} \gtrsim 103.5$ GeV provided by chargino searches at LEP [61]. LEP has also provided lower limits on slepton masses, of which the strongest is $m_{\tilde{e}} \gtrsim 99$ GeV [75]. This is shown by dot-dashed curve in the lower left corner of Fig. 19.

Another important constraint is provided by the LEP lower limit on the

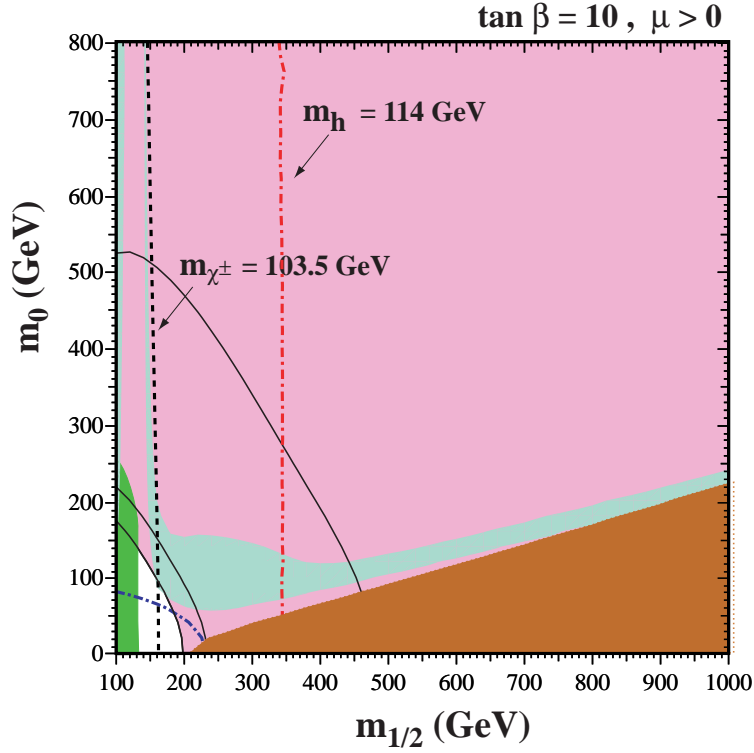

Figure 19. Compilation of phenomenological constraints on the CMSSM for $\tan\beta = 10, \mu > 0$, assuming $A_0 = 0, m_t = 175$ GeV and $m_b(m_b)_{SM}^{\overline{MS}} = 4.25$ GeV [73]. The near-vertical lines are the LEP limits $m_{\chi^\pm} = 103.5$ GeV (dashed and black)[61], and $m_h = 114.1$ GeV (dotted and red)[74]. Also, in the lower left corner we show th $m_{\tilde{e}} = 99$ GeV contour[75]. In the dark (brick red) shaded regions, the LSP is the charged $\tilde{\tau}_1$, so this region is excluded. The light(turquoise) shaded areas are the cosmologically preferred regions with $0.1 \leq \Omega h^2 \leq 0.3$ [76]. The medium (dark green) shaded regions are excluded by $b \to s\gamma$ [77]. The shaded (pink) regions in the upper right regions delineate the $2\sigma$ range of $g_\mu - 2$. The $\pm 1\sigma$ contours are also shown as solid black lines inside the shaded region.

Higgs mass: $m_H > 114.1$ GeV [74]. This holds in the Standard Model, for the lightest Higgs boson $h$ in the general MSSM for $\tan\beta \lesssim 8$, and almost always in the CMSSM for all $\tan\beta$. Since $m_h$ is sensitive to sparticle masses, particularly $m_{\tilde{t}}$, via loop corrections, the Higgs limit also imposes important constraints on the CMSSM parameters, principally $m_{1/2}$ as seen in Fig. 19.

The constraints are evaluated using FeynHiggs[80], which is estimated to have a residual uncertainty of a couple of GeV in $m_h$.

Also shown in Fig. 19 is the constraint imposed by measurements of $b \to s\gamma$ [77]. These agree with the Standard Model, and therefore provide bounds on MSSM particles, such as the chargino and charged Higgs masses, in particular. Typically, the $b \to s\gamma$ constraint is more important for $\mu < 0$, but it is also relevant for $\mu > 0$, particularly when $\tan\beta$ is large. The region excluded by the $b \to s\gamma$ constraint is the dark shaded (green) region to the left of the plot.

The final experimental constraint considered is that due to the measurement of the anomalous magnetic moment of the muon. The BNL E821 experiment reported last year a new measurement of $a_\mu \equiv \frac{1}{2}(g_\mu - 2)$ which deviated by 2.6 standard deviations from the best Standard Model prediction available at that time [81]. The largest contribution to the errors in the comparison with theory was thought to be the statistical error of the experiment, which will soon be significantly reduced, as many more data have already been recorded. However, it has recently been realized that the sign of the most important pseudoscalar-meson pole part of the light-by-light scattering contribution[82] to the Standard Model prediction should be reversed, which reduces the apparent experimental discrepancy to about 1.6 standard deviations.

As many authors have pointed out[83], a discrepancy between theory and the BNL experiment could well be explained by supersymmetry. As seen in Fig. 19, this is particularly easy if $\mu > 0$. The medium (pink) shaded region in the figure is the new $2\sigma$ allowed region: $-6 < \delta a_\mu \times 10^{10} < 58$. With the change in sign of the meson-pole contributions to light-by-light scattering, good consistency is also possible for $\mu < 0$ so long as either $m_{1/2}$ or $m_0$ are taken sufficiently large.

In addition to the coannihilation region discussed above, another mechanism for extending the allowed CMSSM region to large $m_\chi$ is rapid annihilation via a direct-channel pole when $m_\chi \sim \frac{1}{2}m_A$ [84,76]. This may yield a 'funnel' extending to large $m_{1/2}$ and $m_0$ at large $\tan\beta$, as seen in Fig. 20 [73].

### 6.2 Detection

As an aid to the assessment of the prospects for detecting sparticles at different accelerators, benchmark sets of supersymmetric parameters have often been found useful, since they provide a focus for concentrated discussion. A set of proposed post-LEP benchmark scenarios[85] in the CMSSM are illustrated schematically in Fig. 21. They take into account the direct searches for sparticles and Higgs bosons, $b \to s\gamma$ and the preferred cosmological density range. All but one of the benchmark points are consistent with $g_\mu - 2$ at the

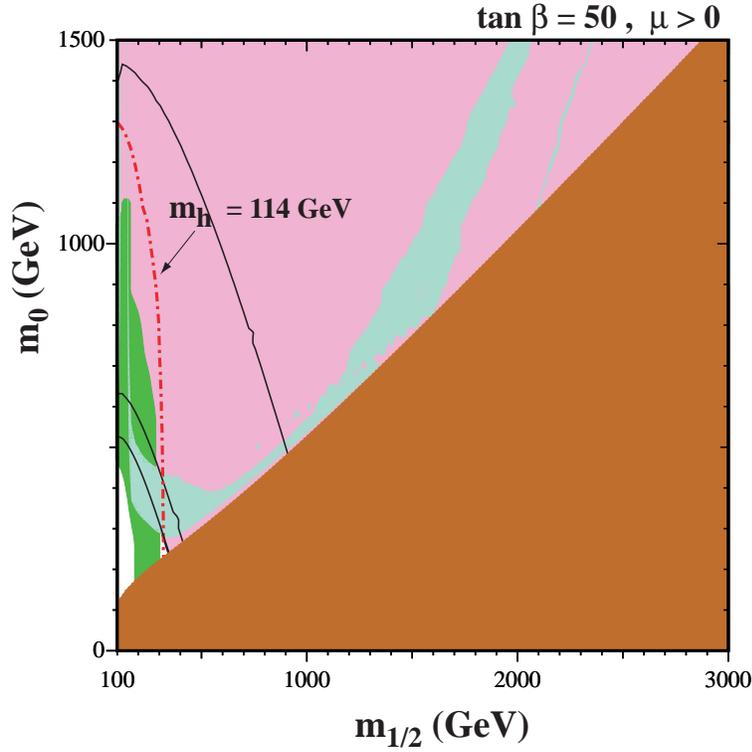

Figure 20. As in Fig. 19 for $\tan\beta = 50$ [73].

$2\,\sigma$ level.

The proposed points were chosen not to provide an 'unbiased' statistical sampling of the CMSSM parameter space, but rather are intended to illustrate the different possibilities that are still allowed by the present constraints[85]. Five of the chosen points are in the 'bulk' region at small $m_{1/2}$ and $m_0$, four are spread along the coannihilation 'tail' at larger $m_{1/2}$ for various values of $\tan\beta$, two are in rapid-annihilation 'funnels' at large $m_{1/2}$ and $m_0$. At large values of $m_0$, (larger than that shown in Figs. 19 and 20, there is another region where the cosmological range is satisfied, namely in the 'focus-point' region[86] along the boundary where electroweak symmetry no longer occurs (shown in Fig. 21 as the shaded region in the upper left corner.) Two points were chosen in the 'focus-point' region at large $m_0$. The proposed points range over the allowed values of $\tan\beta$ between 5 and 50. Prospects for the detection of supersymmetry at future colliders was studied extensively in [85] for these points in particular.

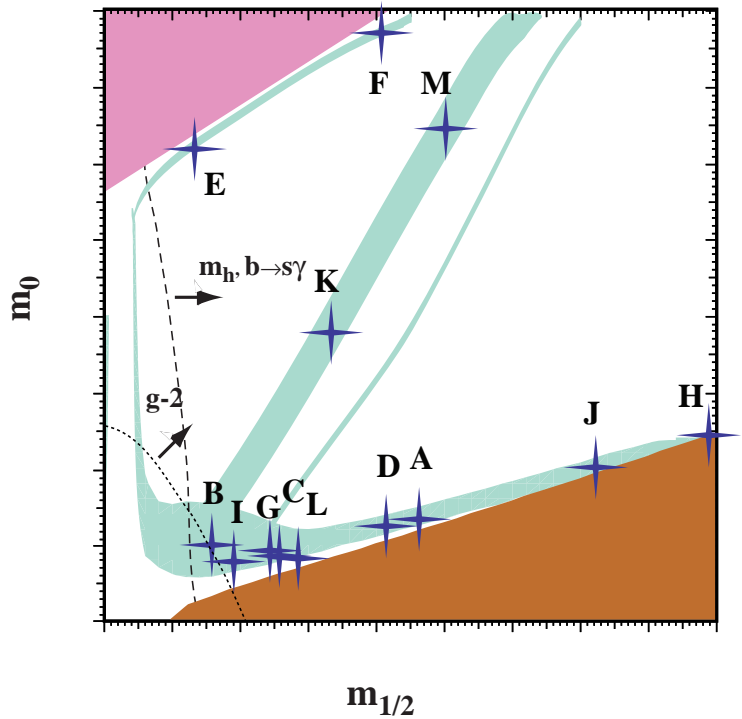

Figure 21. Schematic overview of the CMSSM benchmark points proposed in [85]. The points are intended to illustrate the range of available possibilities. The labels correspond to the approximate positions of the benchmark points in the $(m_{1/2}, m_0)$ plane. They also span values of $\tan\beta$ from 5 to 50 and include points with $\mu < 0$.

Because, the LSP as dark matter is present locally, there are many avenues for pursuing dark matter detection. Here I conclude by showing the prospects for direct detection for the benchmark points discussed above[87]. Fig. 22 shows rates for the elastic spin-independent scattering of supersymmetric relics[87], including the projected sensitivities for CDMS II [88] and CRESST[89] (solid) and GENIUS[90] (dashed). Also shown are the cross sections calculated in the proposed benchmark scenarios discussed in the previous section, which are considerably below the DAMA[91] range ($10^{-5} - 10^{-6}$ pb), but may be within reach of future projects. Indirect searches for supersymmetric dark matter via the products of annihilations in the galactic halo or inside the Sun also have prospects in some of the benchmark scenarios[87].

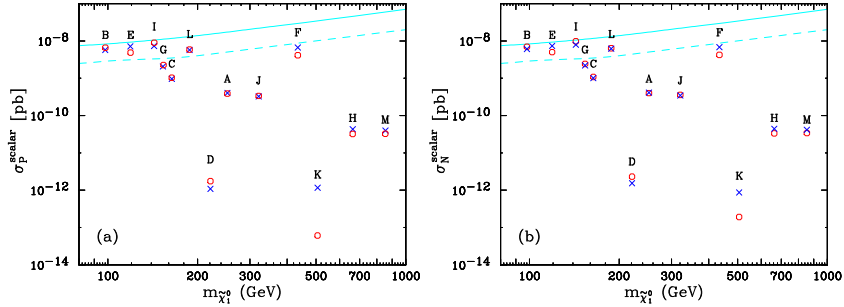

Figure 22. Elastic spin-independent scattering of supersymmetric relics on (a) protons and (b) neutrons calculated in benchmark scenarios[87], compared with the projected sensitivities for CDMS II [88] and CRESST[89] (solid) and GENIUS[90] (dashed). The predictions of our code (blue crosses) and `Neutdriver`[92] (red circles) for neutralino-nucleon scattering are compared. The labels A, B, ...,L correspond to the benchmark points as shown in Fig. 21.

## Acknowledgments

This work was supported in part by DOE grant DE-FG02-94ER40823 at Minnesota.